\documentstyle[referee]{laa}

\begin{document}
\thesaurus{02.14.1; 06.01.2; 02.05.1; 08.09.3}
\title{ A Dynamic Solar Core Model: on the activity-related changes of the neutrino fluxes}

\author{Attila Grandpierre}

\institute{Konkoly Observatory, P.O. ~Box ~67, H--1525,
\newline Budapest, Hungary \\
\newline electronic mail: grandp@ogyalla.konkoly.hu}
\maketitle
\eject

\vskip 8cm

\begin{abstract}
I point out that the energy sources of the Sun may actually involve runaway nuclear reactions as well, developed by the fundamental thermonuclear instability present in stellar energy producing regions. In this paper I consider the conjectures of the derived model for the solar neutrino fluxes in case of a solar core allowed to vary in relation to the surface activity cycle. The observed neutrino 
flux data suggest a solar core possibly varying in time. In the dynamic solar model the solar core involves a "
quasi-static" energy source produced by the quiet core with a lower than standard temperature which may vary in time. Moreover, the solar core involves 
another, dynamic energy source, which also changes in time. The sum of the two different energy sources may produce quasi-constant flux in the SuperKamiokande because it is sensitive to neutral currents, axions and anti-neutrinos, therefore it observes the sum 
of the neutrino flux of two sources which together produce the solar luminosity. A dynamic solar core model is developed to calculate the contributions of the runaway source to the individual neutrino detectors. The results of the dynamic solar core model suggest that since the HOMESTAKE detects mostly the high energy electron neutrinos, therefore the HOMESTAKE data may aniticorrelate with the activity cycle. Activity correlated changes are expected to be present only marginally in the GALLEX and GNO data. The gallium detectors are sensitive mostly to the pp neutrinos, and the changes of the pp neutrinos arising from the SSM-like core is mostly compensated by the high-energy electron neutrinos produced by the hot bubbles of the dynamic energy source. The dynamic solar model suggests that the GALLEX data may show an anti-correlation, while the SuperKamiokande data may show a correlation with the activity cycle. Predictions of the dynamic solar model are presented for the SNO and Borexino experiments which can distinguish between the effects of the MSW mechanism and the consequences of the dynamic solar model. The results of the dynamic solar model are consistent with the present heioseismic measurements and can be checked with future heioseismic measurements as well.


\end{abstract}
%
%************************ SECTION 1
%
\section{Introduction}
The presence of the neutrino problems extends not only to the field of astrophysics (Bahcall, Krastev, Smirnov, 1998), but to the anomalies of the atmospheric neutrinos. 
Apparently, the possible solutions to these unexpected anomalies are not compatible allowing three neutrino flavours (Kayser, 1998).

Nevertheless, the apparent paradox may be resolved when taking into account the effects of the thermonuclear runaways present in stellar cores (Grandpierre, 1996). The neutrinos produced by the runaways may contribute to the events detected in the different neutrino detectors. Estimating the terms arising from the runaways, the results obtained here are at present compatible even with standard neutrinos.
The objections raised against a possible astrophysical solutions to the solar neutrino problems (see e.g. Hata, Bludman, Langacker, 1994, Heeger, Robertson, 1996, Berezinsky et al., 1996, Hata, Langacker, 1997) are valid only if the assumption, that the solar luminosity is supplied exclusively by $pp$ and $CNO$ chains, is fulfilled. Therefore, when this assumption is not 
fulfilled, i.e. a runaway energy source supplies in part the total solar energy production, a more general case has to be considered. In this manuscript I attempt to show that the presence of the runaway energy source, indicated already by first physical principles (Grandpierre, 1977, 1984, 1990, 1996; Zeldovich, Blinnikov, Sakura, 1981) could be described in a mathematically and physically consistent way. The contributions of the runaway source to the neutrino detector data may be determined, allowing also 
solar cycle changes in the neutrino production. The results of Fourier analysis (Haubold, 1997) and wavelet analysis 
(Haubold, 1998) of the new solar neutrino capture rate data for the Homestake experiment revealed periodicities close to 10 
and 4.76 years.

%
%**************************************SECTION 2
%

\section{ Basic equations and the SSM-like approach}
The basic equations of the neutrino fluxes in the standard solar models are the followings (see e.g. Heeger, Robertson, 1996):
\begin{eqnarray}
S_K = a_{K8} \Phi_8
\end{eqnarray}
\begin{eqnarray}
S_C = a_{C1} \Phi_1 + a_{C7} \Phi_7 + a_{C8} \Phi_8
\end{eqnarray}
\begin{eqnarray}
S_G = a_{G1} \Phi_1 + a_{G7} \Phi_7 + a_{G9} \Phi_8 ,
\end{eqnarray}
with a notation similar to that of Heeger and Robertson (1996): the subscripts
i = 1, 7 and 8 refer to $pp + pep$, $Be + CNO$ and $B$ reactions. The $S_j$-s are the observed neutrino fluxes at the 
different neutrino detectors, in dimensionless units, j = K, C, G to the SuperKamiokande, chlorine, and gallium detectors. 
The observed averaged values are $S_K = 2.44$ (SuperKamiokande Collaboration, 1998), $S_C = 2.56$ (Cleveland et al., 1988) and $S_G = 76$ (Kirsten, 1998). $\Phi_i$ are measured in $10^{10} \nu cm^{-2}s^{-1}$.
Similar equations are presented by Castellani et al. (1994), Calabresu et al.
(1996), and Dar and Shaviv (1998) with slightly different parameter values. Using these three detector-equations to determine the individual neutrino fluxes
$\Phi_i$, I derived that
\begin{eqnarray}
\Phi_8 = S_K/a_{K8}
\end{eqnarray}
\begin{eqnarray}
\Phi_1 = (a_{G7}S_C -a_{C7}S_G + S_K/a_{K8}(a_{C7}a_{G8} - a_{G7}a_{C8}))/ a_{G7}a_{C1} - a_{C7}a_{G1}
\end{eqnarray}
and
\begin{eqnarray}
\Phi_7 = (a_{G1}S_C - a_{C1}S_G + S_K/a_{K8}(a_{C1}a_{G8}-a_{G1}a_{C8}))/ a_{G1}a_{C7} - a_{C1}a_{G7}.
\end{eqnarray}
Now let us see how these relations may serve to solve the solar neutrino problems. There are three solar neutrino problems distinguished by Bahcall (1994, 1996, 1997): the first is related to the lower than expected neutrino fluxes, the second to the problem of missing beryllium neutrinos as relative to the boron neutrinos, and the third to the gallium detector data which 
do not allow a positive flux for the beryllium neutrinos in the frame of the standard solar model. It is possible to find a solution to all of the three neutrino problems if we are able to find positive values for all of the neutrino fluxes in the above presented relations. I point out, that the condition of this requirement can be formulated with the following inequality: 

\begin{eqnarray}
S_K < (a_{G1}S_C-a_{C1}S_G)/(a_{C1}a_{G8}/a_{K8}-a_{G1}a_{C8}/a_{K8})
\end{eqnarray}
Numerically,
\begin{eqnarray}
\Phi_7 = 0.4647 S_C -0.0014S_G - 0.5125S_K.
\end{eqnarray}
If we require a physical $\Phi_7>0$, with the numerical values of the detector sensitivity coefficients, this constraint will take the following form:
\begin{eqnarray}
S_K < 0.9024S_C - 0.0027S_G \simeq 2.115.
\end{eqnarray}
In the obtained solutions the total neutrino flux is compatible with the observed solar luminosity $L_{Sun}$, but the reactions involved in the SSM (the $pp$ and $CNO$ chains) do not produce the total solar luminosity. 
The detector rate inequalities (7) or (9) can be fulfilled only if we separate a term from $S_K(0)$,  $S_K(x)$ which
represents the contribution 
of non-pp,CNO neutrinos to the SuperKamiokande measurements (and, possibly, also allow the existence of $S_C(x)$ and $S_G(x)$). The presence of a non-electron neutrino term in the SuperKamiokande is interpreted until now as indication to neutrino oscillations. Nevertheless, thermal runaways are indicated to be present in the solar core that may produce high-energy electron neutrinos, as well as muon and tau neutrinos, since $T>10^{11} K$ is predicted for the hot bubbles (Grandpierre, 1996). Moreover, the explosive reactions have to produce high-energy axions to which also only the SuperKamiokande is sensitive (Raffelt, 1996, Engel et al., 1990). Also, the SuperKamiokande may detect electron anti-neutrinos arising from the hot bubbles. 
This indication suggests a possibility to interpret the neutrino data with standard neutrinos as well.

To determine the $S_i(x)$ terms I introduced the "a priori" knowledge on the pp,CNO chains, namely, their temperature dependence. This is a necessary step to subtract more detailed information from the neutrino detector data. In this way one can derive the temperature in the solar core as seen by the different type of neutrino detectors. 
I note that finding the temperature of the solar core as deduced from the observed neutrino fluxes does not involve the 
introduction of any solar model dependency, since the neutrino fluxes of the SSM pp,CNO reactions depend on temperature only 
through nuclear physics. Instead, it points out the still remaining solar model dependencies of the previous SSM calculations and allowing other types of chains, it removes a hypothetical limitation, and accepting the presence of explosive chains as well, it probably presents a better approach to the actual Sun.

The calculations of the previous section suggested to complete the
SuperKamiokande neutrino-equation with a new term
\begin{eqnarray}
S_K = T^{24.5} \Phi_8(SSM) a_{K8} + S_K(x),
\end{eqnarray}
where $T$ is the dimensionless temperature $T = T(actual)/T(SSM)$.
The one-parameter allowance describes a quiet solar core with a temperature distribution similar to the SSM, therefore it leads to an SSM-like solution of the standard neutrino flux equations (see Grandpierre, 1998). 
An essential point in my calculations is that I have to use the temperature dependence proper in the case when the luminosity is not constrained by the
SSM luminosity constraint, because another type of energy source is also present. The SSM luminosity constraint and the resulting composition and density readjustments, together with the radial extension of the different sources of neutrinos, modify this temperature dependence. The largest effect arises in the temperature dependence of the $pp$ flux: $\Phi_1 \propto T^{-1/2} $ for the SSM luminosity constraint (see the results of the Monte-Carlo simulations of Bahcall, Ulrich, 1988), but $\Phi_1 \propto T^4$ without the SSM luminosity constraint. Inserting the temperature-dependence of the individual neutrino fluxes for the case when the solar luminosity is not constrained by the usual assumption behind the SSM (Turck-Chieze and Lopes, 1993) into the chlorine-equation, we got the temperature dependent chlorine neutrino-equation
\begin{eqnarray}
S_C(T) = a_{C1}T_{C,0}^4 \Phi_1(SSM) + a_{C7}T_{C,0}^{11.5} \Phi_7(SSM) +
a_{C8}T_{C,0}^{24.5} \Phi_8(SSM) +S_C(x)
\end{eqnarray}
Similarly, the temperature-dependent gallium-equation will take the form:
\begin{eqnarray}
S_G(T) = a_{G1}T_{G,0}^4 \Phi_1(SSM)+ a_{G7}T_{G,0}^{11.5} \Phi_7(SSM) +
a_{G8}T_{G,0}^{24.5} \Phi_8(SSM) +S_G(x).
\end{eqnarray}

Now let us first determine the solutions of these equations in the case $S_i(x)=0$. The obtained solutions $T_i$ will be 
relevant to the SSM-like solar core. Now we know that the Sun can have only one central temperature $T$. Therefore, the 
smallest $T_i$-s will be the closer to the actual $T$ of the SSM-like solar core, and the larger $T_i$-s will indicate the terms arising from the runaways. In this way, it is possible to determine the desired quantities $S_i(x)$.

From the observed $S_i$ values, it is easy to obtain $\Phi_1(SSM) = 5.95 \times 10^{10} cm^{-2}s^{-1}$, $\Phi_7(SSM) = 0.594 \times 10^{10} cm^{-2}s^{-1}$ and $\Phi_8(SSM) = 0.000515 \times 10^{10}cm^{-2}s^{-1}$. With these values, the chlorine neutrino-temperature from (11) $T_{Cl} \simeq 0.949 T(SSM)$, the gallium neutrino-temperature is from (12) $T_{Ga} \simeq 0.922 T(SSM)$ and the SuperKamiokande neutrino-temperature is from (10) $T_{SK} \simeq 0.970 T(SSM)$. The neutrino flux equations are highly sensitive to the value of the temperature. Assuming that the actual Sun follows a standard solar model but with a different 
central temperature, the above result shows that the different neutrino detectors see different temperatures. This result 
suggest that the different neutrino detectors show sensitivities different from the one expected from the standard solar model, 
i.e. some reactions produce neutrinos which is not taken into account into the standard solar model, and/or that they are sensitive to different types of non-pp,CNO runaway reactions. Let us explore the consequences of this conjecture.

%
%******************************Section 3
%

\section {Dynamic models of the solar core}

1.) Static core

Obtaining a Ga neutrino-temperature is $T_{Ga} \simeq 0.922 T(SSM)$, this leads to a pp luminosity of the Sun around $L_{pp} \simeq 72 \% L(SSM)$. The remaining part of the solar luminosity should be produced by the hot bubbles, $L_b \simeq 28 \% L(SSM)$. The runaway nuclear reactions proceeding in the bubbles (and possibly in the microinstabilities) should also produce 
neutrinos, 
and this additional neutrino-production, $\Phi_b$ should generate the surplus terms in the chlorine and water Cherenkov detectors as well. At present, I was not able to determine directly, which reactions would proceed in the bubbles, and so it is 
not possible to determine directly the accompanying neutrino production as well. Nevertheless, it is plausible that at 
high temperatures $10^{10-11} K$ (Grandpierre, 1996),  such nuclear reactions occur as at nova-explosions or other types of stellar explosions. Admittedly, these could be rapid hydrogen-burning reactions, explosive CNO cycle, and also nuclear reactions producing heat but not neutrinos, like e.g. the explosive
triple-alpha reaction (Audouze,Truran, Zimmermann, 1973, Dearborn, Timsley, Schramm, 1978). At present, I note that the 
calculated bubble luminosity ($ \simeq 28 \%$) may be easily consistent with the calculated non-pp,CNO neutrino fluxes $\Delta S_{Cl} = S_C(T_{Cl}=0.949) - S_C(T_{Ga}=0.922) \simeq 1.04$, $\Delta S_{Cl}/S_{Cl} \simeq 41 \%$, and $\Delta S_{SK} = S_{SK}(T_{SK}=0.970) - S_{SK}(T_{Ga}=0.922) \simeq 1.74$, $\Delta S_{SK}/S_{SK} \simeq 71 \%$.

The above results are in complete agreement with the conclusion of Hata, Bludman and Langacker (1994), namely: "We conclude 
that at least one of our original assumptions are wrong, either (1) Some mechanism other than the $pp$ and the $CNO$ chains generates the solar luminosity, or the Sun is not in quasi-static equilibrium, (2) The neutrino energy spectrum is distorted by some mechanism such as the MSW effect; (3) Either the Kamiokande or Homestake result is grossly wrong." These conclusions are concretised here to the following statements: (1) a runaway energy source is present in the solar core, and the Sun is not in a 
thermodynamic equilibrium, (2) this runaway source distorts the standard neutrino energy spectrum, and perhaps the MSW effect 
also contributes to the spectrum distortion (3) The Homestake, Gallex and SuperKamiokande results contains a term arising from the non-pp,CNO source, which has the largest contribution to the SuperKamiokande, less to the Homestake, and the smallest to the Gallex. 

The helioseismic measurements are regarded as being in very good agreement with the SSM. However, the interpretation of these 
measurements 
depends on the inversion process, which uses the SSM as its basis. Moreover, the different helioseismic measurements at 
present are contradicting below $0.2 R_{Sun}$ (Corbard et al., 1998).
 
We can pay attention to the fact that the energy produced in the solar core do not necessarily pours into thermal energy, as other, non-thermal forms of energy may also be produced, like e.g. energy of magnetic fields. The production of magnetic fields can significantly compensate the change in the sound speed related to the lower temperature, as the presence of magnetic fields may accelerate the propagation of sound waves with the inclusion of magnetosonic and Alfven magnetohydrodynamical waves.

The continuously present microinstabilities should produce a temperature distribution with a double character, as part of ions may posses higher energies. Their densities may be much lower than the respective ions closer to the standard thermodynamic equilibrium, and so they may affect and compensate the sound speed in a subtle way. Recent calculations of the non-maxwellian character of the energy distribution of particles in the solar core (Degl'Innocenti et al., 1998) indicate that the non-maxwellian character leads to lowering the SSM neutrino fluxes and, at the same time, produces higher central temperatures. This effect may also compensate for the lowering of the sound speed by lowering of central temperature. 

At the same time, an approach specially developed using helioseismic data input instead of the luminosity constraint, the seismic solar model indicates a most likely solar luminosity around $0.8 L_{Sun}$(Shibahashi, Takata, 1996, Figs. 7-10), which leads to a seismological temperature lower than its SSM counterpart, $\Delta T \simeq 6 \%$. On the other hand, as Bludman et al. (1993) pointed out, the production of high energy $^8B$ neutrinos and intermediate energy $^7Be$ neutrinos depends very sensitively on the solar temperature in the innermost $5 \%$ of the Sun's radius. 

Accepting the average value of $R_K = S_{Kam}(obs.)/S_{Kam}(SSM) = 0.474$, this value gives $T_K \simeq 0.97$. With $S_G = 
73.4 SNU$, the derived gallium-temperature will be $T_G \simeq 0.93$. With $S_C = 2.56$ (Cleveland, 1998), $T_C \simeq 
0.95$. The result that $S_G(x)<S_C(x)<S_K(x)$ can arise from the circumstance that the gallium detectors are less sensitive to 
intermediate and 
high-energy neutrinos than the chlorine one, which detects less runaway neutrino than the SuperKamiokande. Therefore, if thermonuclear runaways produce intermediate- and/or high-energy neutrino flux in the Sun, it results a relatively smaller contribution in the gallium detectors than in the chlorine one. Moreover, the SuperKamiokande can detect also runaway muon and tau neutrinos besides the high-energy electron neutrinos, therefore they can contribute with an extra term which would give account why the Kamiokande observes a larger neutrino flux than the Homestake. Therefore, the deduced three temperatures actually indicate that the solar core is actually cooler than the standard one by an amount around $7 \%$. Therefore, the beryllium neutrino flux in the dynamical solar model is estimated as $43 \%$ of its SSM expected value.
The luminosity of the SSM-like core is around $75 \%$ of $L_{Sun}$, therefore the bubble luminosity has to be $25 \% L_{Sun}$. The boron neutrino flux of the SSM like core will be $16.9 \%$ of the SSM value. Therefore, the bubbles has to produce the remaining $\Phi_b = 30.5 \% \Phi_K(SSM)$ of the high energy neutrinos observed by the SuperKamiokande. This requirement may be 
easily satisfied and it may be consistent with the result obtained that the bubble luminosity is $25 \%$ of the solar 
luminosity, too.

The dynamic solar model predicts a beryllium neutrino flux $ \le 43 \%$ of 
the SSM value, corresponding to a temperature of $T(DSM) \le 93 \%$. This estimation offers a prediction for the Borexino neutrino detector $\Phi(Borexino) = \Phi_{Be}(SSM) \times T(DSM)^{11.5} + \Phi(bubbles) \le 43 \% + \Phi(bubbles)$. Regarding the SNO
 detector, I can assume that the neutral currents are produced by the electron neutrinos of the SSM-like core plus all kinds of neutrinos produced by the hot bubbles. Therefore, the prediction of the DSM is $\Phi(SNO) = \Phi(SSM) \times T(DSM)^{24.5} + \Phi(bubbles) \simeq 17 \% + \Phi(bubbles)$. 
These predictions differ significantly from the MSW SSM-values. Therefore, the future observations may definitively decide 
which model describes better the actual Sun, the SSM-based MSW effect or the dynamic solar model. In the interpretation of the 
future measurements it will be important also to take into account the possible dependence of the neutrino fluxes on the solar cycle.

2.) Around activity maximum

Similarly, we can apply the equations given to derive the temperatures as seen by the different neutrino detectors in relation to the phases of solar activity. Around solar activity maximum the Kamiokande reported no significant deviancies from the averaged neutrino flux, therefore I can take $R_K(max) = 0.474$ which leads to $T_K(max) = 0.97$ again. With the data of Cleveland et 
al. 
(1998), neutrino fluxes were measured in two solar activity maximum period, in 1980 the result was $17.2 \%$ and in 1989 around $42.5 \%$, which compares to the reported average value of $47.8 \%$. Since the average absolute flux is $2.56 SNU$, this refers to an expected flux of $5.36 SNU$. These values leads to $S_C(max) \simeq 1.60 SNU$. Also, the Gallex collaboration did not report about activity related changes in their observed neutrino data, therefore $S_G(max) = 76 SNU$ can be used. Solving the 
neutrino flux equations for an assumed SSM-like solar core, the resulting temperatures will be $T_C(max) \simeq 93 \%$ and 
$T_G(max) \simeq 0.922$. 

The obtained results, $T_K(max) \simeq 0.97$, $T_C(max) \simeq 0.93$, $T_G(max) \simeq 0.92$, are consistent with the 
assumption that in the solar maximum the Gallex and the Homestake detect only the neutrinos from the SSM-like solar core, which has a temperature around $8 \%$ lower than in the SSM, or that in the solar maximum the neutrinos produced by the hot bubbles contribute mostly to the SuperKamiokande data. This results may be regarded as well fitting to the main point of the paper, namely that the neutrino flux produced by the hot bubbles produce muon and tau neutrinos, axions and anti-neutrinos, to which only the SuperKamiokande is sensitive.  

3.) Around solar activity minimum

Using a value $R_K(min) = 0.474$ for the minimum of the solar activity, the Kamiokande temperature will be $T_K(min) \simeq 0.97$. With the data presented in Cleveland et al. (1998), in the periods around solar minimum the Homestake measured $0.823 \ counts \ day^{-1} $ around 1977, $0.636 \ counts \ day^{-1}$ around 1987, and $0.634 \ counts \ day^{-1} $ around 1997. These 
values average to $0.299 \ counts \ day^{-1}$, suggesting an $S_C(min) \simeq 3.737$ SNU. With this $S_C(min)$ the neutrino 
flux equations leads to $T_C(min) \simeq 0.97$. Now the Gallex results marginally indicates larger than average counts around 1995-1997, as reported by the Gallex-IV measurements of $117 \pm 20 \ SNU$. This value leads to a $T_G(min) \simeq 0.99$, i.e. an anti-correlation with the solar cycle. For a temperature of $T_G(min) \simeq 0.97$ the $S_G(min)$ would be $\simeq 96 \ SNU$.

The results obtained above suggest that around solar minimum all the neutrino detector data are consistent with a uniform temperature $T_K(min) 
= T_C(min) = T_G(min) = 0.97$. In this case the results would suggest that all the neutrino detectors observe only the SSM-like solar core and not the neutrinos arising from the hot bubbles of the thermonuclear runaways. In the dynamic solar model there 
is a quick and direct contact between the solar surface and the solar core. In the dynamic solar model the transit time 
scale of the hot blobs from he solar core to the surface is estimated to be around one day (Grandpierre, 1996), therefore, the 
absence of the surface sunspots may indicate the simultaneous absence (or negligible role) of runaways in the core.  
Therefore, the result that in solar minimum no bubble neutrino flux are observed in each of the neutrino detectors, is consistent with the fact that in solar minimum there are no (or very few) sunspots observed at the solar surface.

%
%**************************************SECTION 4
%
\section{Discussion and Conclusions}

The calculated solutions of the neutrino flux equations are consistent with the data of the neutrino detectors. I have shown that introducing the runaway energy source, it is possible to resolve the apparent contradiction between the different neutrino detectors even assuming standard neutrinos. Moreover, the results presented here suggest that the physical neutrino problems of the atmospheric neutrinos may be consistent with the solution of the solar neutrino problems even without introducing sterile neutrinos. 

Considering the hypothetical activity-related changes of the solar neutrino fluxes, I found that the twofold energy source of the Sun produces different contributions in the different neutrino detectors. Apparently, it is the SuperKamiokande that is the most sensitive to the runaway processes. The contribution of the runaway neutrinos and the neutrinos of the standard-like 
quiet solar core runs in anti-correlation to each other. Therefore, their effects may largely compensate each other in the 
SuperKamiokande data. Nevertheless, it is indicated that intermediate and high-energy neutrinos may produce a slight correlation with the solar activity in the SuperKamiokande data since they correlate more closely with the runaway neutrino fluxes than with the neutrinos of the SSM-like solar core. They give a $64 \%$ of the total counts observed in the SuperKamiokande, therefore, the total flux 
may slightly correlate with the solar cycle. On the contrary, since the Homestake do not see the runaways, except the intermediate and high-energy electron neutrinos produced by the hot bubbles, its data may anti-correlate with the solar activity. 
Moreover, the dynamic solar model suggests that GALLEX data may anti-correlate with the solar cycle as well since it is more 
sensitive to the low-energy neutrinos arising from the proton-proton cycle, although it is also sensitive to the intermediate and high-energy electron neutrinos produced by the hot bubbles.

Now, obtaining indications of possible correlations between the solar neutrino fluxes and activity parameters, I can have a short look to the data whether they show or not such changes in their finer details. Such a marginal change may be indicated in the Figure 3. of Fukuda et al. (1996). In this figure, the maximum value is detected just in 1991, at solar maximum, 
consistently with the results obtained here. Moreover, its value as read from that figure seems to be $68 \%$ of the value 
expected from the SSM. In 1995, in solar minimum, the lowest value, $34 \%$ is detected, again consistently with the interpretation we reached. Later on, the SuperKamiokande started to work and measured a value of $2.44 \times 10^6 cm^{-2}s^{-1}$ for the boron neutrino flux. Assuming that the values in 1995 and 1996-1997 did not differ significantly, as it is a period of the solar minimum, the 
two observation can be taken as equal, i.e. the $34 \%$ is equal with $2.44 \times 10^6 cm^{-2}s^{-1}$. This method gives for the $68 \%$ value a boron flux of $4.88 \times 10^6 cm^{-2}s^{-1}$. Now Bahcall, Basu and Pinsonneault (1998) developed an improved standard solar model, with significantly lower $^7Be(p, \gamma)^8B$ cross sections,  $5.15 \ cm^{-2}s^{-1}$ instead of the previo
us $6.6 \times 10^6 cm^{-2}s^{-1}$ of Bahcall and Pinsonneault (1995). With this improved value the $4.88 \times 10^6 cm^{-2}s^{-1}$ leads to a $\Phi_k(min) \simeq 95 \% \ \Phi_K(SSM)$! This means that actually even the SuperKamiokande data may contain some, yet not noticed correlation tendency with the solar cycle. These indications make the future neutrino detector data more interesting to a possible solar cycle relation analysis.

The dynamic solar model has a definite suggestion that below 0.10 solar radius the standard solar model is to be replaced by a significantly cooler and possibly varying core. These predictions can be checked with future helioseismic observations. Helioseismology is not able to tell us the temperature in this deepermost central region. 
On the other hand, the presence of the thermonuclear micro-instabilities causes a significant departure from the thermal equilibrium and changes the Maxwell-Boltzmann distribution of the plasma particles. 
It is shown that such modification leads to increase the temperature of the solar core, which can compensate the non-standard cooling (Kaniadakis, Lavagno, Quarati, 1996) and so the simple dynamic solar model can be easily consistent with the helioseismic results as well.

The indicated presence of a runaway energy source in the solar core - if it will be confirmed - will have a huge significance in our understanding of the Sun, the stars, and the neutrinos. This subtle and compact phenomena turns the Sun from a simple gaseous mass being in hydrostatic balance to a complex and dynamic system being far from the thermodynamic equilibrium. This complex, dynamic Sun ceases to be a closed system, because its energy production is partly regulated by tiny outer influences like planetary tides. This subtle dynamics is possibly related to stellar activity and variability. Modifying the participation of the MSW effect in the solar neutrino problem, the dynamic energy source has a role in the physics of neutrino mass and oscillation.
An achievement of the suggested dynamic solar model is that it may help to solve the physical and astrophysical neutrino problems without the introduction of sterile neutrinos, and, possibly, it may improve the bad fit of the MSW effect (Bahcall. Krastev, Smirnov, 1998).

%
%**************************Section 6
%
\section{Acknowledgements}
The work is supported by the Hungarian Scientific Research Foundation
OTKA under No. T 014224.

%
%******************************References
%
\eject
%\section{References}


\begin{thebibliography}{}
\bibitem{}
Audouze, J., Truran, J. and Zimmermann, B. A. 1973, Astrophys. J. 184, 493
\bibitem{}
Bahcall, J. N., Basu, S. and Pinsonneault, M. H. 1998, Phys. Lett. B433, 1
\bibitem{}
Bahcall, J. N., Krastev, P. I. and Smirnov, A. Yu. 1998, hep-ph/9807216
\bibitem{}
Bahcall, J. N. and Ulrich, R. 1988, Rev. Mod. Phys. 60, 97
\bibitem{}
Berezinsky, V., Fiorentini, G. and Lissia, M. 1996, Phys. Lett. B365, 185
\bibitem{}
Bludman, S. A., Hata, N., Kennedy, D. C., and Langacker, P. O. 1993, Phys. Rev. D49, 2220, hep-ph/9207213 
\bibitem{}
Calabresu, E., Fiorentini, G., Lissia, M. and Ricci, B. 1995, hep-ph/9511286
\bibitem{}
Castellani, V., Degl'Innnocenti, S., Fiorentini, G., Lissia, M. and Ricci, B. 
1994, Phys. Lett. B324, 425
\bibitem{}
Cleveland, B. T., Daily, T., Davis, R., Distel, J. R., Lande, K., Lee, C. K., and Wildenhain, P. S. 1998, ApJ 496, 505
\bibitem{}
Corbard, T., Di Mauro, M. P., Sekli, T., and the GOLF team, 1998, preprint ESA SP-418, Obs. Astrophys. Catania preprint 16/1998
\bibitem{}
Dar, A. and Shaviv, G. 1998, astro-ph/9808098
\bibitem{}
Dearborn, D., Timsley, B. M. and Schramm, D. N. 1978, Astrophys. J. 223, 557
\bibitem{}
Engel, J., Seckel, D. and Hayes, A. C. 1990. Phys. Rev. Lett. 65, 960
\bibitem{}
Fukuda, Y. et al., 1996, Phys. Rev. Lett. 77, 1683
\bibitem{}
Fukuda, Y. et al., The SuperKamiokande Collaboration, 1998, hep-ph/9807003
\bibitem{}
Grandpierre, A. 1977, University Doctoral Thesis, Polytechnic University, Budapest
\bibitem{}
Grandpierre, A. 1990, Sol. Phys. 128, 3
\bibitem{}
Grandpierre, A. 1984, in "Theoretical Problems in Stellar Stability and Oscillation", eds. Noels, A. and Gabriel, M., 48 
\bibitem{}
Grandpierre, A. 1996, Astron. Astrophys. 308, 199
\bibitem{}
Grandpierre, A. 1998, subm. to Phys. Rev. D., astro-ph/9808349
\bibitem{}
Hata, N., Bludman, S., and Langacker, P. 1994, Phys. Rev. D49, 3622
\bibitem{}
Hata, N. and Langacker, P. 1997, Phys. Rev. D56, 6107, hep-ph/9705339
\bibitem{}
Haubold, H. J. 1997, Nucl. Phys. A621, 341c
\bibitem{}
Haubold, H. J. 1998, astro-ph/9803136 
\bibitem{}
Heeger, K. M. and Robertson, R. G. H. 1996, Phys. Rev. Lett. 77, 3720
\bibitem{}
Kayser, B. 1998, The European Phys. J. C3, 1, http://pdg.lbl.gov/
\bibitem{}
Kirsten, T. A. 1998, Progress of Nuclear and Particle Physics, Vol. 40, to appear
\bibitem{}
Raffelt, G. 1997, astro-ph/9707268 
\bibitem{}
Shibahashi, H. and Takata, M. 1996, Publ. Astron. Soc. Japan 48, 377
\bibitem {}
The Super-Kamiokande Collaboration, Fukuda et al., 1998, hep-ph/9807003
\bibitem{}
Turck-Chieze, S. and Lopes, L. 1993, ApJ 408, 347
\bibitem{}
Zeldovich, Ya. B., Blinnikov, S. I., and Sakura, N. I. 1981, The physical basis of stellar structure and evolution, Izd. Moskovskovo Univ., Moskva, 1981, in Russian

\end{thebibliography}
\end{document}